\documentclass[10pt]{iopart}
\begin{document}
\jl{03}

\title[The rheology of field-responsive suspensions]
{The rheology of field-responsive suspensions}

\author{J.M. Rub\'{\i}\dag\  and J.M.G. Vilar\ddag}

\address{\dag\ Departament de F\'{\i}sica Fonamental,
Universitat de Barcelona, Diagonal 647, E-08028 
Barcelona, Spain }

\address{\ddag\ Departments of Physics and Molecular Biology,
Princeton University,
Princeton, New Jersey 08544 }

\begin{abstract}

We present some aspects of the rheology of field-responsive suspensions, a
class of field-responsive liquid matter systems possessing the ability to
undergo significant changes in their strength upon application of an
external field. Both the single-particle and the many-particle domains are
discussed. In the former, consideration of the full nonlinear dynamics of
the particles leads to an anomalous behavior of the viscosity whereas in
the latter the most salient feature is the formation of chains and fractal
structures. We indicate how to deal with the rheology at moderately
concentrations leaving open the problem at higher concentrations for which
the complexity of the emergent structures strongly limits the knowledge
of their dynamics.

\end{abstract}




\section{Introduction}

The term field-responsive liquid matter embraces a class of
soft-condensed-matter systems which exhibit peculiar behavior upon
application of an external field. Of particular interest is the family of
two-phase systems in which one of the phases is active whereas the other
is practically inactive to the effect of the field. The active phase
consists of single dipolar particles which at low concentrations remain
randomly dispersed in the liquid phase whereas at larger concentrations
have a tendency to aggregate forming chains or more complex branched
structures. These structures are embedded in the liquid phase which may be
a simple liquid, a polymer, or a liquid crystal, to name just a few.

As occurs in dispersions of neutral colloidal particles or polymer
solutions, the objects of the suspended phase interact with their fluid
surroundings thus modifying the flow of this fluid. Unlike those systems,
the interactions of the particles with the fluid are assisted by the
external field, moreover, for a given concentration, structures which for
neutral particles would be stable may evolve in time giving rise to
interesting collective phenomena. The consequence of those features is
that some static and dynamic properties of the system as the overall
magnetization or the viscosity may vary substantially as a consequence of
the external field or the dipolar interactions among particles. This
fact constitutes one of the most important characteristics of these
systems, which has been used in many technological applications.
Electro- and magneto-rheological
fluids~\cite{xxx2} and
ferrofluids~\cite{xxx2,Rosenweig2,xxx4,Odenbach,Blums}
belong to that class of systems whose static and
dynamic properties have been subject of great interest in the last years.

Our purpose in this paper is to present several relevant aspects of
the rheology of field-responsive suspensions. We will discuss the
single-particle and the many-particle regimes. For dilute suspensions,
the case of constant field has been extensively studied and shows a
monotonous increasing of the viscosity when the field increases. For
time-dependent fields this behavior breaks down and even in the linear
domain the viscosity may decrease upon increasing the field. This
phenomenon leading to a diminution of the total viscosity of the
suspension has been found recently and has been referred to as the
´negative´ viscosity effect~\cite{Gazeau,Rosenweig1,Perez1}.
Interesting nonlinear effects may arise by the
coupling of the dynamics of the magnetic moment, described by the
Landau-Gilbert equation, to the one for the orientation of the particle.
The dynamics in this regime exhibits a rich phenomenology whose main
macroscopic implications is the breaking of the monotonous behavior of
the viscosity~\cite{Saiz}. The main feature in the moderately concentrated
regime is the formation of chains and branched structures.
Chains appear at low temperatures, for which Brownian effects
are irrelevant, or high dipolar interactions, which occurs when the
particles are induced dipoles (electro- and magneto-rheological fluids).
They are also formed in one-dimensional adsorption processes
in the presence of dipolar interactions~\cite{Pastor2}.
In the opposite limit, including the case of nanosized ferromagnetic
particles or ferrofluids, more compact aggregates may be found.

Kinetic models have been proposed to analyze the dynamics of 
chains of induced dipoles and its implication in the rheological behavior.
The structures emerging at high temperatures are sometimes hierarchical.
The rheology in this regime is more complex and modeling of the dynamics
needs to be developed.

The paper has been organized as follows. In section 2 we analyze the
macroscopic implications of the rotational dynamics of the particles
by establishing a relationship between the viscosity associated to the rotation
of the particle, or rotational viscosity, and its dynamics. This formula is
applied to several cases already addressed in the literature which
in this framework
can be
treated in a very simple and systematic way. Section 3 is devoted to present
some well-known
examples belonging to the single-particle domain. In section 4 we
analyze the nonlinear domain and discuss very briefly the appearance of a
dynamical transition breaking the monotonous nature of the rotational
viscosity as a function of the field. In section 5 we deal with the
rheology of the moderated concentrated phase. Finally in the discussion
section we summarize the main results obtained in the regimes addressed
previously and point out some methodological aspects useful in the
treatment of these systems.

\section{From mesoscopic to macroscopic}

To illustrate the implications of the mesoscopic dynamics of the suspended
particles in the macroscopic behavior of the system we  consider first
the case in which the field-responsive fluid consists of a dilute
suspension of spherical dipolar particles in a simple liquid under the
influence of an external magnetic field which may be constant or
time-dependent. The dipole moments remain rigidly attached to the
particles being parallel to their orientations.
The dynamics of both magnetic moment and orientation of the
particle are given by
\begin{equation}
\label{elR}
\frac{d \hat R}{dt}=\vec \Omega \times \hat R \;,
\end{equation}
where $\hat R$ is the unit vector along the dipole moment and $\vec
\Omega$ is the angular velocity of the particle. 
A very common situation
arises when inertial effects can be neglected. In
this case, the angular velocity of the particle is completely determined
by the balance of magnetic, hydrodynamic, and Brownian torques.
It is given by
\begin{equation}
\label{elBT}
\vec \Omega =
\vec \omega_0+\frac{1}{\xi_r}\vec m \times \vec H
+\frac{1}{\xi_r}\hat R \times \vec F_B(t) \,,
\end{equation}
Here $\vec m =
m_0 \hat R$ is the dipole moment, and $\vec H$ is the applied field.  The
hydrodynamic torque arises from the angular velocity difference between
the particle and the fluid with vorticity $2\omega_0$; its strength also
depends on the rotational friction coefficient, $\xi_r$. Brownian effects
are taken into account through a random force, $\vec F_B(t)$, with zero
mean and correlation function $\left<\vec F_B(t)\vec
F_B(t')\right>=2\xi_rk_BT{\vec{\vec{1}}}\delta(t-t')$, where $k_B$,
$T$, and ${\vec{\vec{1}}}$ are the Boltzmann constant, the absolute
temperature, and the unit tensor, respectively.

Previous equations describe the rotational dynamics of the particle inside
the fluid, but do not give information about
the effects of this mesoscopic motion in the macroscopic flow. In
particular, the presence of colloidal particles in a liquid modifies the
macroscopic properties of the fluid by changing the transport
coefficients. Several methods have been developed to compute the
dependence of the transport coefficients upon the presence of these
particles. Here, however, we follow an alternative approach which
directly relates the mesoscopic dynamics of the particle to its
macroscopic effects. To this purpose we consider the power dissipated in
the fluid by the motion of the particle.  Since the rotational motion
is assisted by the external field, its angular velocity has
not necessarily be adapted to the local angular velocity of the fluid,
i.e. to the local vorticity. This fact originates additional dissipation
in the system resulting from the torque exerted by the particle on
the fluid. Therefore, the power dissipated per unit of volume due to the
presence of the particles is
\begin{equation}\label{eq:g2}
\sigma_{mes} = n\xi_r
(\vec{\omega_0} - \vec{\left<\Omega\right>})\cdot\vec{\omega_0}\;,
\end{equation}
where $n=N/V$ is the concentration of colloidal particles and $<.>$ stands
for average over thermal noise. The rotational viscosity, $\eta_r$,
accounts for the change in the total viscosity of the system due to the
rotational degrees of freedom of the particles. From a macroscopic point
of view, the power dissipated by this change in the viscosity~\cite{Mazur}
is given by
\begin{equation}
\sigma_{mac} = \eta_r({2\omega_0})^2\;.
\end{equation}
The relationship between mesoscopic and macroscopic properties follows
straightforwardly since previous equations are merely two different
expressions for the same quantity, i.e., $\sigma_{mac} = \sigma_{mes}$. In
this way, we obtain
\begin{equation}
\eta_r= {1 \over 4}n\xi_r
\left(1 - {\left<\Omega\right> / \omega_0} \right)\;,
\end{equation}
which leads to the same result as in Ref.~\cite{Perez1} when the rotational
friction coefficient is replaced by its explicit expression for a sphere,
$\xi_r=8\pi\eta_0\,a^3$, with $\eta_0$ and $a$ being the viscosity of the
carrier fluid and the radius of the particle, respectively. Notice,
however, that our expression is not only restricted to spheres, as those
of Ref.~\cite{Gazeau}, but it can also be applied to any other type
of particles by
only considering the appropriate expression for the rotational friction
coefficient.  From this expression the rotational viscosity follows from
the knowledge of the dynamics of the particle in the host medium. It may
reach positive or negative values depending on the ratio between the mean
angular velocity of the particle and the local vorticity.

The methodology introduced previously, far from being
specific, can be applied to a wide variety of situations, including
different types of particles and flows.
Let us outline here its applicability to a rod-like colloid in an
elongational flow 
defined through the velocity field $\vec v_s= \vec{\vec{\kappa}}\cdot
\vec r$, with
$
\vec{\vec{\kappa}}= \beta\left(3\,\hat e_x \hat e_x - \vec{\vec{1}}\right)\;
$
and $\beta$ representing the velocity gradient and the elongational rate,
respectively.
The dynamics of the particle follows in a 
similar way as in the previous situation, but now with the hydrodynamic
torque given by
\begin{equation}
\vec T^H=-\xi_r[\vec\Omega-\hat R \times (\vec{\vec{\kappa}}\cdot \hat R)]\;.
\end{equation}
The power dissipated can be expressed as both
\begin{equation}
\label{disirod}
\sigma_{mes}= n\left<{\vec{\vec {\xi}}}:
({\vec{\vec{\kappa}}}\cdot \hat R
-\vec \Omega \times  \hat R)
({\vec{\vec{\kappa}}}\cdot \hat R)\right>
\end{equation}
and
\begin{equation}
\sigma_{mac}=2\eta_e \vec{\vec{\kappa}}:\vec{\vec{\kappa}}=
12\eta_e \beta^2\;,
\end{equation}
where $
{\vec{\vec {\xi}}}=\xi_{||}\hat R \hat R
+\xi_\perp({\vec{\vec 1}}-\hat R \hat R)
$
is the friction tensor~\cite{Doi,Saluena}.
Therefore, the elongational viscosity is given by
\begin{equation}
\eta_e={\sigma_{mes}\over 12\beta^2} \;.
\end{equation}
Notice that in contrast with the previous situation,
the elongational viscosity depends on correlations
of $\vec \Omega$ and $\hat R$. Vorticity effects can also be considered
in this case. Their contribution to the
rotational viscosity is obtained in a completely analogous way as
previously done.

\section{Some representative examples in the single-particle domain }

Our purpose in this section is to present some illustrative examples which
have already been discussed in the literature. Here, however, we follow
the method proposed in the previous section
from which much of the essential information
obtained from more complicated procedures can be recovered.

The simplest case addressed is the one in which the
suspended dipoles are rigid and relax in a noiseless environment in the
presence of a constant magnetic field. The dynamics of the magnetic moment
is given by Eqs. (\ref{elR}) and (\ref{elBT}) where now
the random force, $\vec F_B(t)$, has to be omitted.
To better illustrate the essentials of the
phenomenon the explicit situation we consider consists in
an applied magnetic field along the $x-$direction,
$\vec H=H_0 \,\hat e_x$, and a vorticity perpendicular to it along the
$y-$direction, $\vec\omega_0=\omega_0 \,\hat e_y$.

The equilibrium orientation of the particle for the previous situation is
given by
\begin{equation}
\hat{{R}}_s=\left\{
\begin{array}{ll}
\sqrt{1-(1/\alpha)^2}\;\hat{{x}}-\; (1 /\alpha)\hat{{z}}
&\;\;\mbox{if}\;\alpha \ge1 \\
\pm \sqrt{1-\alpha^2}\;\hat{{y}}
-\;\alpha \hat{{z}}
&\;\;\mbox{if}\;\alpha < 1 \; ,
\end{array}
\right. 
\end{equation}
where $\alpha={m_{0}H / \xi _{r}\omega _{0}}$ is the ratio between the
magnetic and hydrodynamic torques. The expression for the rotational
viscosity readily follows after substituting the value for the orientation
of the particle in the expression for the angular velocity resulting in
\begin{equation}
\eta_r=\left\{
\begin{array}{ll}
{1 \over 4} n\xi_r
&\;\;\mbox{if}\;\alpha \ge1 \\
{1 \over 4} n\xi_r\alpha^2
&\;\;\mbox{if}\;\alpha <1 \;.
\end{array}
\right. 
\end{equation}
Notice that the rotational viscosity increases quadratically with the
intensity of the applied field until it reaches a saturation value at
fields which are strong enough to prevent particle motion.

Let us now look just at the opposite situation, in which Brownian effects
dominate the dynamics. Taking averages in Eq. (\ref{elBT}), for the
component of the angular velocity parallel to the vorticity we obtain
\begin{equation}
\left<\Omega_y \right> =
\omega_0+\frac{1}{\xi_r}m_0H_0\left<R_z\right>
\,,
\end{equation}
whereas from Eq. (\ref{elR}),
\begin{equation}
\label{estaz}
\frac{d}{dt}\left<R_z\right>=
\omega_0\left<R_x\right>
-\frac{m_0H_0}{\xi_r}\left<R_zR_x^2\right>
-\frac{k_BT}{\xi_r}\left<R_z\right>
\end{equation}
\begin{equation}
\label{estax}
\frac{d}{dt}\left<R_x\right>=
-\omega_0\left<R_z\right>
+\frac{m_0H_0}{\xi_r}\left<1-R_x^2\right>
-\frac{k_BT}{\xi_r}\left<R_x\right>
\end{equation}

From the previous equations we can easily obtain the viscosity for some
representative cases in the stationary state
($d\vec{\left<R\right>}/dt=0$). For instance, in the high field limit we 
have $R_x\sim1$. Therefore,
\begin{equation}
\left<R_z\right>=\frac{\xi_r\omega_0}{m_0H_0+k_BT}\;\;\mbox{and}\;\;
\eta_r=\frac{n\xi_r}{4+4k_BT/m_0H_0}\;.
\end{equation}
In contrast, for high noise or low field
we have $m_0H_0\ll k_BT$, which leads to
$
\left<R_x\right>=\frac{k_BT}{\xi_r\omega_0}\left<R_z\right>\,.
$
Moreover, since the distribution of orientations is almost uniform, i.e.
$\left<R_x^2\right>\sim1/3$,
\begin{equation}
\left<R_z\right>=
\frac{2m_0H_0/\xi_r\omega_0}{3+3(k_BT/\xi_r\omega_0)^2}
\end{equation}
and
\begin{equation}
\eta_r=\frac{1}{24}n\xi_r
\left[\frac{(m_0H_0)^2}{(\xi_r\omega_0)^2+(k_BT)^2}\right]
\;.
\end{equation}

As a common feature, in all these examples in which the magnetic field is
constant, the rotational viscosity is a monotonous function of the field
reaching a saturation limit at high values of the field. This behavior
have been observed experimentally~\cite{MacTague}.

Let us now consider the situation in which the field is oscillating in
time, $\vec H = H_0e^{i\omega t}\hat e_x$. For the sake of simplicity
we assume, as previously, that the
amplitude of the field is sufficiently small.
In this case the mean angular velocity is given by
\begin{equation}
\left<\Omega_y \right> =
\omega_0+\frac{1}{\xi_r}m_0
\overline{\Re(H_0e^{i\omega t})\Re(\left<R_z\right>_\omega)}
\,,
\end{equation}
where $\vec{\left< R \right>} = \vec{\left< R \right>_\omega} e^{i\omega t}$
and the overline stands for time average and $\Re$ for the real part of
its argument. The solution of the previous equations follows
straightforwardly by only realizing that since $d
\vec{\left<R\right>}/dt=i\omega\vec{\left<R\right>}$ the resulting
equations are then formally identical to Eqs. (\ref{estaz}) and
(\ref{estax}) when replacing $k_BT$ by $k_BT+i\xi_r\omega$. Taking the
real part of the viscosity obtained in that way leads to
\begin{equation}
\eta_r=
n\frac{m_0^2H_0^2}{24\xi_r}
\left\{\frac{\omega_0^2-\omega^2+(k_BT/\xi_r)^2}
{\left[(\omega_0^2-\omega^2+(k_BT/\xi_r)^2\right]^2
+2(k_BT/\xi_r)^2\omega^2}\right\}\;,
\end{equation}
which corresponds to the increase of the viscosity due to the presence of
the oscillating field. As a remarkable feature, it is worth to point out
that this correction to the viscosity can take negative as well as
positive values. Negative values in this quantity means that the mean
angular velocity of the particle is higher than that of the fluid.
Therefore, energy of the oscillating field is transformed into kinetic
energy that contributes to diminish the effective viscosity of the fluid.
Notice that
for this situation to happen, $\vec{\left<R_z\right>}$ and $H_x$ must be
anticorrelated, which never can happen when
$d\vec{\left<R_z\right>}/dt=0$.

Particles of different shape can also be included in this framework.
For instance, in the case of a suspension of rod-like
particles, the elongational viscosity follows from the dissipated power
[Eq. (\ref{disirod})]. This quantity can be decomposed into its
elastic, $\sigma^E$, and viscous, $\sigma^V$, contributions,
i.e., $\sigma_{mes} = \sigma^E + \sigma^V$:
\begin{eqnarray}
\sigma^E &=&n\xi_\perp\left<\left[\hat R \times
(\vec{\vec{\kappa}}\cdot \hat R)-\vec\Omega\right]
\cdot \hat R \times (\vec{\vec{\kappa}}\cdot \hat R)\right>\\
\sigma^V&=&
n\xi_{||}\left<(\hat R \cdot (\vec{\vec{\kappa}}\cdot \hat R))^2\right>
\;,
\end{eqnarray}
Therefore, for an elongational flow and a magnetic field applied
along the $x$-direction we obtain
\begin{eqnarray}
\sigma^E &=& 3n\beta\left[{k_BT}\left<(3R_x^2-1)\right>
- m_0H_0\left<R_x(1-R_x^2)\right>\right]\\
\sigma^V&=&n\beta^2\xi_{||}\left<(3R_x^2-1)^2\right>\;,
\end{eqnarray}
where the averages involved can be computed from the probability
distribution of $R_x$~\cite{Doi,Saluena},
\begin{equation}
P(R_x) \propto
\exp{\left({m_0H_0\over K_BT}R_x+{\beta\xi_r\over K_BT}(3R_x^2-1)\right)}\;.
\end{equation}
From previous equations then we can easily obtain the elongational viscosity.
This and other types of viscosities 
have been computed in Ref.~\cite{Saluena} following a different
approach. They have been found to depend also on the
magnetic field and the aspect ratio of the particles.  
Notice that here the dependence on the form of the particles enters through
the explicit values of $\xi_{\perp}$ and $\xi_{||}$.

\section{Rheology in the  nonlinear regime}

When the magnetic moment of the particle is not necessarily rigidly attached,
it is oriented along an intermediate direction between those of the magnetic
field and the easy axis of magnetization.  The two relaxation
mechanisms can be identified through the energy of the particles which for
uniaxial crystals reads
\begin{equation}
\label{a1}
\label{primera}
U=-\vec{m}\cdot\vec{H}-K_aV_p(\hat{n} \cdot \hat{R})^2\;,
\end{equation}
where $K_a$ is an effective constant incorporating contributions from
crystalline and shape anisotropies, $V_p$ is the volume of one of these
spheres, and $\hat{n}$ is the unit vector along the direction of the
easy axis of magnetization. The expression for the energy includes the case of
rigid dipole, in which the vectors $\hat R$ and $\hat n$ are parallel, and
the case of soft dipoles
in sufficiently high magnetic fields, in which the magnetic moment
orients itself in the direction of
the field very quickly, then the particle rotates towards the stationary
orientation where $\hat R$ and $\hat n$ are parallel. In
this case the energy of the particle reduces to the energy of anisotropy.

The dynamics of the magnetic moment is governed by the Landau-Gilbert
equation. When the ferromagnetic particle is rotating itself
with angular velocity $\vec\Omega$ it expresses as
\begin{equation}
\label{a2}
\frac{d\hat{R}}{dt} =
- h \hat{R}\times \frac{\partial U}{\partial\hat{R}}\times \hat{R}
+(\vec\omega_L+\vec{\Omega})\times \hat{R}, 
\end{equation}
Here, ${\vec\omega}_L=-g{\partial U /\partial \hat{R}}$ is the Larmor
frequency and $h$ and $g$ are constants. Additionally, the orientation of
the easy axis of magnetization, $\hat n$, evolves according to the
rigid rotor equation [Eq. (\ref{elR})]
whereas the angular velocity follows from the balance of
torques [Eq. (\ref{elBT})].

The dynamics of the particle has been solved analytically for the case of
low vorticity and magnetic field~\cite{Shliomis1}, belonging
to the validity domain of linear response theory. In this sense the
rotational viscosity can be computed through a Green-Kubo
formula~\cite{Miguel} and as
in the case of the rigid dipole exhibits a monotonous behavior when
represented as a function of the field. Additionally, from that formalism
one can compute the relaxation time of the magnetic moment towards the
field for different substances~\cite{Miguel} arriving at good agreement with
birefringence experiments~\cite{Bacri}.

The nonlinear dynamics described by Eqs. (\ref{a1}), (\ref{a2}),
and the non-rigid dipole counterpart
of Eqs. (\ref{elR}) and (\ref{elBT})
reveals the existence of a rich phenomenology. As discussed previously, in the
linear regime deviations from the monotonous behavior of the rotational
viscosity have only been found for time-dependent magnetic fields. In the
case of constant magnetic field the viscosity always increases with the
field reaching a saturation limit~\cite{MacTague,Shliomis2}.
In the nonlinear domain the orientation
of the particle may explore states which otherwise would not be accessible
opening the possibility for the appearance of dynamical transitions among
different states. Upon increasing of the vorticity the system passes from
a state in which the particle behaves essentially as a rigid dipole to
another in which the orientation of the particle relaxes towards the
magnetic moment which
in turns undergoes an oscillatory motion around the imposed field.
The macroscopic implications of this dynamics is important and manifests in
the rotational viscosity. Anomalous behavior of this coefficient has been
found even at constant magnetic field~\cite{Saiz}. In the first state
the viscosity
increases with the field as usual but in the second it decreases. During
this dynamical transition the system can exhibit hysteresis.

\section{Rheology in the moderately concentrated phase}

When the concentration of particles increases, dipolar interactions become
to be important and aggregation process takes place. To avoid this process
the magnetic particles are coated with a surfactant which introduces a
steric repulsion. In the regime of low concentrations the interactions
balance each other out and the suspension is stable. Beyond that regime,
particles aggregate, the system then loses its original nature evolving
towards more stable configurations. This process is in the most general
situation a dynamical process in which the structures may change in time
by growth or fragmentation of some parts and may in general be controlled
by noise, the presence of an external field (constant or time-dependent)
and the action of an imposed velocity field. When the individual particles
assemble into chains, the length has been found to depend on time
through a scaling law whose predicted dynamic exponent agrees with
experimental
results. In the absence of field dipolar particles selforganize
hierarchically and fractal structures are formed.  Their shapes depend on
the competition of dipolar forces and thermal agitation and can be
characterized by means of the fractal dimension of the aggregates.
The probabilities of growth, $P_g$, and splitting, $P_s$, of the
structure are related
through~\cite{Pastor1}
\begin{equation}
{P_g \over P_s} \sim e^{1/T_r} \;,
\end{equation}
where $T_r$ is a reduced temperature comparing thermal and dipolar energies.
This expression may intuitively account for the
appearance of different structures. In the limit $T_r \rightarrow 0$,
since $P_s \sim 0$, chains are the most likely structure.
In the opposite limit, $P_s \sim P_g$, spherical aggregates are
formed. The fractal dimension
of the aggregates ranges from $1.1$, at low temperatures, to $1.7$
which corresponds to the limit of pure diffusion-limited-aggregation,
occurring at high temperatures. Consequently, the condition of clusters
mutually opaque: $D_1 + D_2 > 2$, where $D_1$ and $D_2$ are the
fractal dimensions of
two interacting clusters, holds and they may behave in solution as spheres
with an effective radius.  The appearance of fractal structures predicted
in two dimensions~\cite{Pastor1} has been corroborated by experiments about
aggregation
of magnetic particles in Langmuir monolayers~\cite{Lefebure}.
 
The analysis of the rheological properties at higher concentrations
constitutes an open problem which needs the implementation of new
theoretical approaches. In the regime in which particles selforganize
forming chains, kinetic models have been proposed in which the length of
the chains may vary in time due to growth and fragmentation processes.
The dynamics of the orientation of the chain is in essence
the one of a nonlinear oscillator. As an example, for a chain
of induced dipoles rotating under the influence of both
field and shear flow the angle of the chain with the direction
of the field evolves according with~\cite{Martin},
\begin{equation}
\dot\theta+\omega_0\sin(2\theta)=\dot\gamma\cos^2\theta\;,
\end{equation}
where $\omega_0$ is a characteristic frequency and $\dot\gamma$
the shear rate. The dynamics of other elongated structures can be 
formulated in a similar way~\cite{Halsey}.
These models have predicted the behavior of the viscosity as a function of
the shear rate in particular situations when the field is constant.
The rheology is very sensitive to the form of the underlying structure
which tacitly implies that its characterization must be very precise.
The case of time-dependent field is more complex an new nonlinear
rheological behavior have been observed~\cite{Negita}.
Modeling at higher concentration, for which chains interact and
branched structures are formed, constitutes a challenge for future
developments.  

The methods proposed based upon Fokker-Planck dynamics, dealing
essentially with single particles, could in principle be generalized by
considering the presence of dipolar interactions among particles as a new
ingredient.  There are only a few results in this context. Extension of
the Fokker-Planck dynamics, developed for the case of single particles, to
the case of dipolar interactions has been carried out to analyze
the dynamics of a moderately concentrated suspension of ferromagnetic
particles~\cite{Zubarev}. Fokker-Planck dynamics has also been treated
perturbatively to calculate the viscosity of a suspension of chains
of ferromagnetic particles in an elongational flow~\cite{Miguel2}.
As we have seen previously, rheological properties depend
on the dynamics of the emergent structures and therefore the knowledge of
those properties is highly limited when the dynamic of the different
entities integrating the solid phase is unknown.

\section{Discussion}

Along this paper we have discussed rheological properties of
field-responsive fluids. The main characteristics of these
systems regarding the dynamics, is that it is assisted by the external
field and it
is influenced by dipolar interactions, which are the dominant ones.  The
single-particle regime is much more complex than in the case of neutral
colloids and therefore its study has an intrinsic interest. The dynamics
of a single particle is in general nonlinear and its description requires
the knowledge of the evolution of two vectors: the magnetic moment and the
orientation of the particle, whose dynamics are coupled. The linear domain
is well-known and can be analyzed by continuum theories or linear response
theory, the classical results for the viscosity as a function of the field
have been obtained in this situation. The nonlinear dynamics has not been
explored in depth and deserves more attention. The analysis of some
particular situations belonging to that domain makes it evident the
presence of new phenomenology, sometimes counterintuitive. The
many-particle domain has as a main feature the formation of structures
mainly chains or branched structures which may in general evolve in time.
Information about rheological properties at moderate concentrations when
chains are the emergent structures can be obtained from kinetic models. In
the more concentrated regime in which the very notion of chain breaks down
new approaches have to be implemented.

The methodology to analyze the dynamics of these systems is quite similar
to the one used in the study of other soft-condense-matter systems. At the
phenomenological level, continuum theories may provide information in the
dilute regime. Scaling laws can be proposed for example to analyze the
length of chains or the form of fractal aggregates. Fokker-Planck dynamics
provide always an useful framework to analyze dynamical properties as for
example the viscosity. This method has limitations when the system is
complex, which is is case concerning us. Even in the linear regime one
obtains hierarchies of equations for the moments which may provide
information after decouplings. When dipolar interactions are considered,
Fokker-Planck dynamics can only be solved perturbatively for particular
situations. Contrasting with this methodology, in the
formalism we have proposed the viscosity is given in terms of the angular
velocity of the particle which in turn may easily be computed via
simulations. This procedure may provide a useful framework from which
rheological properties of these systems could be analyzed.
Numerical simulations on mesoscopic and macroscopic scales~\cite{Bossis,Hess}
constitute an increasingly valuable adjunct to theoretical and experimental
studies.

In this paper we have dealt mainly with ferrofluids, a class of
field-responsive systems in which the particles are single domain magnetic
particles with permanent dipole moments. The methods we have proposed and
the problems we have outlined also hold or have a counterpart in the case in
which the particle bear induced dipoles as occurs in electro- and
magneto-rheological fluids or even in suspensions of latex particles in a
ferrofluid, the so-called magnetic holes whose nonlinear dynamics has been
analyzed in assemblies of dimers~\cite{Helgensen}.

\section*{Acknowledgments}

We want to acknowledge R. Pastor-Satorras, M.C. Miguel
and A. P\'erez-Madrid for useful discussions. This work has been supported
by the DGES of the Spanish Government under
grant PB95-0881 and by the INCO-COPERNICUS program of the European
Commission under contract IC15-CT96-0719. J.M.G.V. acknowledges financial
support from the Ministerio de Educaci\'on y Cultura (Spain).

\end{document}